\documentclass[final,5p,times,twocolumn]{elsarticle}

\usepackage{graphicx}
\usepackage{amssymb}
\usepackage{amsmath}
\usepackage{amsfonts}
\usepackage{graphicx}
\usepackage{longtable}
\newcommand{\be}{\begin{equation}}
\newcommand{\ee}{\end{equation}}

\newcommand{\ba}[1]{\left(\begin{array}{#1}}
\newcommand{\ea}{\end{array}\right)}

\journal{Results in Physics}

\begin{document}

\begin{frontmatter}



\title{Is composite noise necessary for sudden death of entanglement?}


\author[a]{Yashodamma K.O} 
\author[a,b]{Sudha}
\ead{arss@rediffmail.com}

\address[a]{Department of Physics, Kuvempu University, Shankaraghatta, Shimoga-577 451, India.} 
\address[b]{Inspire Institute Inc., Alexandria, Virginia, 22303, USA.}
\begin{abstract}
The finite time disentanglement or entanglement sudden death, when only one part of the composite system is subjected to a {\emph single} noise, is 
examined. While it is shown that entanglement sudden death can occur when a part of the entangled mixed state is subjected to either amplitude noise or phase noise, local action of either of them does not cause entanglement sudden death in pure entangled states. In contrast, depolarizing noise is shown to have an ability to cause sudden death of entanglement even in pure entangled states,  when only one part of the state is exposed to it. The result is illustrated through the action of different noisy environments individually on a single qubit of the so-called X class of states and an arbitrary two-qubit pure state.  
\end{abstract}

\end{frontmatter}


\section{Introduction}
\label{sec1}
Quantum entanglement~\cite{n1, n2}, a special kind of correlation that exists in composite quantum systems has been of great relevance both for its importance in foundational issues of quantum mechanics~\cite{n1,n2} and for its use as resource in quantum computation~\cite{n3, n4}. The fragility of entangled quantum systems~\cite{n5,n6} to the environment surrounding them is the main cause of concern for realizing their technological aspirations.   
Decoherence refers to the process through which superposition of quantum states are irreversibly transformed into statistical mixtures due to coupling of quantum system with surrounding environment \cite{n5,n6,n7,n8,n10}. Eventhough  it is not possible to fully characterize the environment, different models have been proposed. Early stage disentanglement or Entanglement Sudden Death (ESD)~\cite{n13a,n13,n14a,n14} is a consequence of the exposure of entangled systems to noisy environment or channels. 

In \cite{n13}, Yu and Eberly elucidated the concept of finite time disentanglement by examining how two entangled qubits individually interacting with pure vacuum noise lose their entanglement in a finite time. Further examination of ESD has revealed that this early stage disentanglement is an illustration of nonadditivity of individual coherence decay rates \cite{n14}. This inevitable loss of entanglement has gained lot of attention in the quantum information community~\cite{n13a,n13,n14a,n14,sda,sdb,n16,n16a,sd1,sd2,sd3,aj,sd4,sd5,sd6,sd7,sk,sd8,sd9,sd10,new1} and efforts are being made to find ways in which ESD can be avoided~\cite{lf,arp1,arp2,new2,new3,new4}.   

It is to be noted here that the studies on finite time disentanglement in bipartite systems~\cite{n13a,n13,n14a,n14,sda,sdb,n16,n16a,sd1,sd2,sd3,sd4,sd5,sd6,sd7,sd8,sd9,sd10} concentrated on situations in which either both the subsystems are coupled individually to one or more noisy environments or one of the subsystems is coupled to more than one noise. While it can be interpreted that nonadditivity of individual coherence decay rates is manifested in the occurence of ESD in such cases~\cite{n14}, a question of interest is whether exponential decay of entanglement always results when one of the subsystems alone is coupled to a noisy environment leaving the other subsystem entirely free of noise. Except in Ref. \cite{aj,sk}, where sudden death of entanglement is reported in qubit-qutrit states when only one of them is either subjected to dephasing~\cite{aj} or depolarizing noise~\cite{sk}, not much attention is paid to the finite time disentanglement due to a single local noise.  We wish to take up this issue and examine the disentanglement dynamics of bipartite states when one of the subsystems is kept free of noise and the other is subjected to a {\emph {single}} noisy environment. 

The article is organized as under: Sec. \ref{sec1} contains introductory remarks. In Sec. \ref{sec2}, we have examined the effect of amplitude, phase and depolarizng noises each acting locally on a qubit of two-qubit X states and two-qubit pure states.  Two special states of the class of X states, the isotropic, Werner states are analyzed under the local action of these three noises in Sec. \ref{sec3}. Concluding remarks are given in Sec. \ref{sec4} 

\section{Sudden death of entanglement in a class of X states}
\label{sec2}
We have considered the class of two qubit states of the form,
\be
\label{ye}
\rho=
\begin{bmatrix}
a & 0 & 0 & 0  \\  0 &  b & z & 0 \\ 0 & z^* & c & 0 \\ 0 & 0 & 0 & d
\end{bmatrix}
\ee
where $ a+b+c+d=1$,  $ z = x+iy$.  This is a subset of the so-called X class of states~\cite{n13, n14, n16a} and it  encompasses  both pure states such as the all important Bell states and mixed states such as Werner and isotropic states. 

It is readily seen that concurence~\cite{n17,n18}, a well known measure of entanglement for two-qubit states, of the above state is of the form
\be
\label{con} 
C= 2\,\mbox{Max} \left[ 0,\left| z\right|-\sqrt{ad} \right ]. 
\ee
The noisy channels we have considered here are amplitude noise, phase noise and depolarizing noise. It is well known that the effect of each of these noises on a quantum system can be characterized through the corresponding Kraus operators~\cite{n3,n13,n14,yucl}. 

\subsection{Amplitude Noise:}
The process such as spontaneous emission of a photon is characterized by the quantum operation called amplitude damping or amplitude noise. It gives the right description for energy dissipation from a quantum system.  
The Kraus operators for a single qubit amplitude noise are given by \cite{n13}
\be
E_{0}=
\begin{bmatrix}
\eta & 0  \\  0 &  1
\end{bmatrix}; \ 
E_{1}=
\begin{bmatrix}
0 & 0  \\  \sqrt{1-\eta^{2}} &  0
\end{bmatrix}
\ee
where $\eta = e^{-\frac{\Gamma_a t}{2}}$ and $\Gamma_a$ denotes the longitudinal decay rate~\cite{n14}. 

When only the first qubit is subjected to amplitude noise leaving the second qubit noise free, the corresponding Kraus operators are given by
\begin{eqnarray*}
K_{1a}&=&
\begin{bmatrix}
\eta & 0  \\  0 &  1
\end{bmatrix}\otimes
\begin{bmatrix}
1 & 0  \\  0 &  1
\end{bmatrix} \nonumber \\ 
& & \\
K_{2a}&=&
\begin{bmatrix}
0 & 0  \\  \sqrt{1-\eta^{2}} &  0
\end{bmatrix}\otimes
\begin{bmatrix}
1 & 0  \\  0 &  1
\end{bmatrix} \nonumber 
\end{eqnarray*}
The time evolved density matrix with amplitude noise acting on first qubit of state Eq. (\ref{ye}) is obtained as,
\begin{eqnarray*}
\label{y1}
\rho_a(t)&=& K_{1a} \rho K_{1a}^\dagger + K_{2a}\rho K_{2a}^\dagger \nonumber \\
&=& 
\begin{bmatrix}
a \eta^2 & 0 & 0 & 0  \\  0 &  b\eta^2 & z \eta  & 0 \\ 0 & z^* \eta & c+a- a\eta^2 & 0 \\ 0 & 0 & 0 & d+b-b\eta^2
\end{bmatrix}
\end{eqnarray*}
It can be readily seen that the state $\rho_A(t)$ retains its X-form and hence its concurrence can be calculated using Eq. (\ref{con}). We get, 
\be
\label{y2}
C_{a}= 2\, \mbox{Max} \left[ 0,\eta\left(\left| z\right|-\sqrt{a[b+d-b \eta^2]} \right)\right].
\ee
It can be seen that $ C_{a} =0 $ when $\vert z \vert - \sqrt{a(b+d-b\eta^2)}\leq 0$  
or $ t \geq \frac{1}{\Gamma_a} \ln \left [ \frac{ab}{a(b+d)- \vert z \vert^2 } \right]$. 
Thus, depending on the parameters of the initial density matrix, one can either observe sudden death of entanglement or not. This feature is illustrated in Fig. 1.  

In particular, when the state under consideration is a Bell state $\vert \Psi^{\pm}\rangle=\frac{1}{\sqrt{2}}\left(\vert 01 \rangle\pm \vert 10 \rangle \right)$, 
the denisty parameters turn out to be $a=d=0$ and $b=c=\vert z \vert=1/2$. The concurrence of the state, under the action of amplitude noise on its first qubit, turns out be $C_a= \mbox{Max} [0, e^{-\frac{\Gamma_a\,t}{2}}]$ indicative of exponential decay of entanglement.  

\begin{figure}[h]
\includegraphics* [width=2.2in,keepaspectratio]{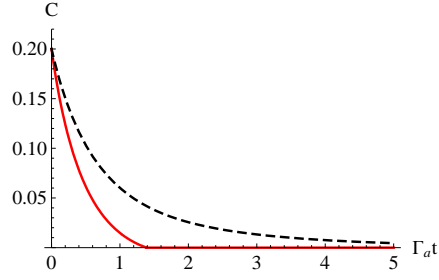}
\caption{ When the density matrix parameters are chosen to be $a=0.1, b=0.4, c=0.4, d=0.1$ and $ \vert z\vert^2 = 0.04$ (solid line) the entanglement of the state vanishes when $\Gamma_a t= 1.386$. But when the density matrix parameters are chosen to be $a=0.1,\  b=0.2,\  c=0.6,\  d=0.1$ and $ \vert z\vert^2=0.04$ there is an exponential decay of entanglement (dashed line) contrary to the previous case.}
\end{figure}

\subsection{Phase Noise} 
Phase damping or phase noise is a uniquely quantum mechanical noise that describes the loss of quantum information without loss of energy. 
The Kraus operators for single qubit phase noise are given by~\cite{yucl}
\be
K_{0}=
\begin{bmatrix}
1 & 0  \\  0 &  \gamma
\end{bmatrix};\ \ 
K_{1}=
\begin{bmatrix}
0 & 0  \\  0 &  \sqrt{1-\gamma^{2}}
\end{bmatrix}
\ee where $\gamma =e^{-\frac{\Gamma_p t}{2}}$ and $\Gamma_p$ denotes transverse decay rate~\cite{n14}.

The Kraus operators for the action of phase noise on the first qubit alone being $K_{1p}=K_{1}\otimes I_2$, $K_{2p}=K_{2}\otimes I_2$ ($I_2$ denotes the $2\times 2$ identity matrix), we have 
\begin{eqnarray}
\label{y3}
\rho_{p}(t)&=& K_{1p} \rho K_{1p}^\dagger + K_{2p}\rho K_{2p}^\dagger \nonumber \\
&=& \begin{bmatrix}
a  & 0 & 0 & 0  \\  0 & b & z \gamma & 0 \\ 0 & z^* \gamma & c & 0 \\ 0 & 0 & 0 & d
\end{bmatrix} 
\end{eqnarray}
We can observe that only the off diagonal terms are affected but the time evolved density matrix retains its X-form. On evaluating the concurrence for this state, we get
\be
\label{y4}
C_{p}= 2\, \mbox{Max} \left[ 0,\gamma \vert z\vert-\sqrt{ad}\right]
\ee
Notice that $ C_{p}= 0 $ when $ \left( \gamma\vert z\vert-\sqrt{ad} \right)\leq  0 $ or $ t \geq \frac{1}{\Gamma_p} \ln\left[\frac{\vert z\vert^2}{ad}\right]$.
As in the case of amplitude noise, depending on the initial parameters of the density matrix, one can either observe sudden dealth of entanglement or not. The illustration of the result is done for some specified values of $a,\ b,\ c,\ d$ and $ \left| z\right|^2 $ in Fig. 2. 

It can be observed even here that the Bell state $\vert \Psi^{\pm}\rangle=\frac{1}{\sqrt{2}}\left(\vert 01 \rangle\pm \vert 10 \rangle \right)$  
does not show any finite time disentanglement as $C_p=\mbox{Max}[0,\,e^{-\frac{\Gamma_p t}{2}}]$ indicates exponential decay of entanglement. 
\begin{figure}[h]
\includegraphics* [width=2.2in,keepaspectratio]{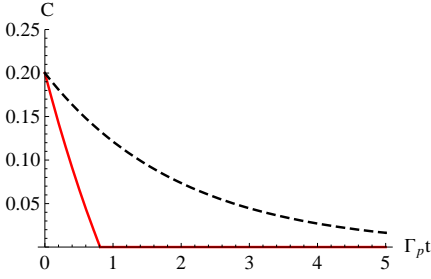}
\caption{While the density matrix with parameters $a=0.2, b=0.3, c=0.3, d=0.2 $ and $ \left| z\right|^2 = 0.09 $ give rise to sudden death of entanglement (solid line) due to single-qubit phase noise, there is no such finite time disentanglement (dashed line) in the density matrix $\rho$ with $ a=0.5,\  b=0.1,\  c=0.4,\  d=0 $ and $ \left| z\right|^2=0.01 $}
\end{figure}

\subsection{The Depolarizing noise} 
A quantum noise that converts a qubit into a completely mixed state with probability $p$ and leaves it untouched with probability $1-p$ is the depolarizing noise~\cite{n3}. The Kraus operators for depolarizing noise are given by~\cite{n3} 
\begin{eqnarray}
D_{1}&=& \sqrt{1-p}
\ba{rr}
1 & 0  \\  0 &  1
\ea
; \ \ 
D_{2}= \sqrt{\frac{p}{3}}
\ba{rr}
0 & 1  \\  1 & 0
\ea; \nonumber \\
D_{3}&=&\sqrt{\frac{p}{3}}
\ba{rr}
0 & i  \\  -i &  0
\ea; \ \
D_{4}=\sqrt{\frac{p}{3}}
\ba{rr} 
1 & 0  \\  0 &  -1
\ea 
\end{eqnarray}
with $p=1-e^{-\Gamma_d\,t/2}$.  
 
To observe the effect of depolarizing noise on entanglement when it acts only on the first qubit of a two-qubit system, we have to consider the Kraus operators,
\begin{eqnarray*}
K_{1d}&=&D_{1}\otimes I_2,\ \ K_{2d}=D_{2}\otimes I_2, \\ 
K_{3d}&=&D_{3}\otimes I_2,\ \ K_{4d}=D_{4}\otimes I_2.
\end{eqnarray*}

The time evolved density matrix with depolarizing noise acting on first qubit alone is given by,
\begin{eqnarray}
\label{y5}
& & \rho_d{(t)}=\sum_{i=1}^4 \, K_{id}\,\rho\,K_{id}^\dagger  \\
&=&\ba{cccc} 
a+\frac{2p(c-a)}{3}  & 0 & 0 & 0  \\  0 & b+\frac{2p(d-b)}{3} & \frac{(3-4p) z}{3}  & 0 \\ 0 & \frac{(3-4p) z^* }{3} & c+\frac{2p(a-c)}{3} & 0 \\ 0 & 0 &0 & d+\frac{2p(b-d)}{3} \nonumber 
\ea 
\end{eqnarray}
As the resultant density matrix has retained its X-form even under the depolarizing noise, we can readily obtain the concurrence of the state. It is given by 
\begin{small}
\be
\label{y6}
C_{d}= \frac{2}{3} \,\mbox{Max} \left[0,\,(3-4p)\vert z\vert-\sqrt{[3a+2p(c-a)][3d+2p(b-d)]}\right]
\ee 
\end{small}
In Fig. 3, we have illustrated two situations where amplitude and phase noise do not cause sudden death of entanglement while 
depolarizing noise causes it. 
\begin{figure}[h]
\includegraphics* [width=2.2in,keepaspectratio]{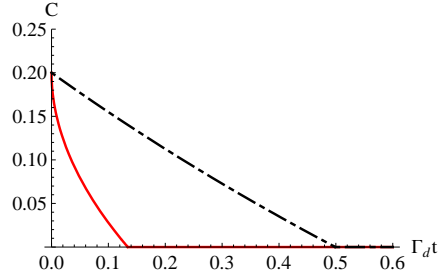}
\caption{While the initial density matrix parameters are chosen to be $a=0.5, b=0.1, c=0.4, d=0$  and $ \vert  z\vert^2 = 0.01 $ in one case(solid line), they are $ a=0.1, b=0.2, c=0.6, d=0.1 $ and $ \vert z\vert^2=0.04$ in the other (dot-dashed line).  Sudden death of entanglement, due to single-qubit depolarizing noise, is observed in both the cases.}
\end{figure}  

In contrast to the case of amplitude and phase noise, depolarizing noise can cause finite time disentanglement in the Bell states $\vert \Psi^{\pm}\rangle$. In fact, $C_d= \mbox{Max} \left[0,\, 1-2p\right]$ for $\vert \Psi^{\pm}\rangle$ indicating finite time disentanglement when $p\geq 1/2$. It is of importance to notice here that depolarizing noise causes sudden death of entanglement even in an arbitrary pure state of the form 
\be
\label{pure}
\vert\phi\rangle=\sqrt{a} \vert 00  \rangle+\sqrt{b} e^{if}\vert 01  \rangle+
\sqrt{c} e^{ig} \vert 10  \rangle+\sqrt{d} e^{ih} \vert 11  \rangle; \ \ a+b+c+d=1.
\ee     
with initial concurrence 
\be
C_{ini}^P=\mbox{Max}\left[0,\,2\sqrt{ad+bc-2\sqrt{a\,b\,c\,d}\cos(f+g-h)}\right]
\ee
Denoting the concurrence of the state $\vert\phi\rangle$ when amplitude, phase, depolarizing noise act individually on its first qubit by $C_a^P$, $C_p^P$ and $C_d^P$ respectively, we have on explicit evaluation, 
\begin{small}
\be
C_a^P=\mbox{Max}\left[0,\,2e^{-\frac{\Gamma_a\, t}{2}}\sqrt{ad+bc-2\sqrt{a\,b\,c\,d}\cos(f+g-h)}\right] \nonumber
\ee
\be
C_p^P=\mbox{Max}\left[0,\,2e^{-\frac{\Gamma_p\, t}{2}}\sqrt{ad+bc-2\sqrt{a\,b\,c\,d}\cos(f+g-h)}\right] 
\ee
\be 
C_d^P=\mbox{Max}\left[0,\,2(2e^{-\frac{\Gamma_d\, t}{2}}-1)\sqrt{ad+bc-2\sqrt{a\,b\,c\,d}\cos(f+g-h)}\right] \nonumber 
\ee
\end{small}
It can be readily seen through the expressions for $C_a^P$, $C_p^P$
that both amplitude and phase noises lead to exponential decay of entanglement in an initially entangled state $\vert \phi\rangle$ ( $C_{ini}^P\neq 0$). But the expression for $C_d^P$ indicates that depolarizing noise can cause sudden death of entanglement in $\vert \phi\rangle$  when  
$2e^{-\Gamma_d\, t/2}-1\leq 0$ or when  $t\geq \frac{2}{\Gamma_d}\ln 2$.  It is to be noted here that the time at which initial entanglement vanishes in pure states $\vert \phi \rangle$ does not depend on the state parameters but only on the decay rate $\Gamma_d$ of the depolarizing noise.  This feature is illustrated in Fig. 4.
\begin{figure}[h]
\includegraphics* [width=2.2in,keepaspectratio]{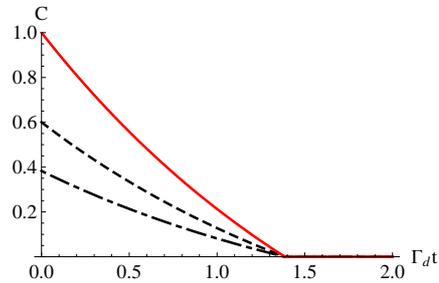}
\caption{Finite time disentanglement caused by single qubit depolarizing noise in the state $\vert\phi\rangle=\sqrt{a} \vert 00  \rangle+\sqrt{b} e^{if}\vert 01  \rangle+
\sqrt{c} e^{ig} \vert 10  \rangle+\sqrt{d} e^{ih} \vert 11  \rangle$ with (i) $a=d=1/8;\ b=c=3/8; f=g=h=\pi/4$ (dashed line), (ii) 
$a=d=1/4;\ b=c=1/4; f=g=h=\pi/4$ (dot-dashed line) (iii) $a=d=1/2;\ b=c=0; f=g=h=0$ (solid line) corresponding to the Bell State $\vert \Phi^{+}\rangle=\frac{1}{\sqrt{2}}\left(\vert 00 \rangle+\vert 11 \rangle \right)$. Notice that for all states $\vert \phi \rangle$ (with $C_{ini}^P\neq 0$), entanglement vanishes at the same time.} 
\end{figure} 

Due to the ability of depolarizing noise to kill the initial entanglement of pure as well as mixed states in a finite time, we can readily conclude that sudden death of entanglement is a more possible happening in the case of depolarizing noise than in the cases of amplitude or phase noise. 

The family of X states considered in Eq.(\ref{ye}) include mixed quantum states of importance 
such as Werner states and isotropic states. In what follows, we will show that each of these states can undergo sudden death of entanglement under the action of a single noise on a single qubit.  
 
\section{Isotropic and Werner states}
\label{sec3} 
\noindent{\bf Isotropic State}: The isotropic states are $d^2$ dimensional bipartite states that are convex mixtures of a maximally entangled state with a maximally mixed state \cite{jac12}.  They are conveniently expressed in the form~\cite{jac12a}
\be
\label{iso}
\rho_{\rm{iso}} = \frac{1-x}{d^2-1} \left[ I - \vert \psi \rangle  \langle \psi \vert \right]+ x \vert \psi \rangle \langle \psi \vert 
\ee 
where $0 \leq x \leq 1$ and $\vert \psi \rangle$ is any  maximally entangled state of dimension $d^2$. Notice that $\rho_{\rm {iso}}$ is entangled when the parameter $x$ is greater than $\frac{1}{d}$. Isotropic states possess the property of invariance under unitary transformations of the form $U\otimes U^\ast$.   

On choosing $\vert \psi \rangle$ to be the singlet state $\frac{\vert 01 \rangle-\vert 10 \rangle}{\sqrt{2}}$, the two-qubit isotropic state is given by  
\be
\label{y22}
\rho = 
 \ba{cccc}
\frac{1-x}{3} & 0 & 0 & 0 \\  0 & \frac{2x+1}{6}  &  \frac{4x-1}{6}&  0 \\ 
0  &\frac{4x-1}{6} & \frac{2x+1}{6} & 0  \\ 0 & 0 & 0 & \frac{1-x}{3}
\ea.  \ee 
The concurrence of the above state is readily seen to be  
$C_i = \mbox{Max}\,  [0,\,2x-1] $ which implies $ C_i > 0$ for all $x > \frac{1}{2} $. 
An interaction of the state's first qubit alone with amplitude noise leads to 
\be
\label{y23}
\rho_{ia} = 
\ba{cccc}
\frac{(1-x)\eta^2}{3} & 0 & 0 & 0 \\  0 & \frac{(2x+1)\eta^2}{6}  &  \frac{(4x-1)\eta}{6}  &  0 \\ 
0 &   \frac{(4x-1)\eta}{6} & \frac{(3+2(x-1))\eta^2}{6} & 0  \\ 0 & 0 & 0 & \frac{3-(1+2x)\eta^2}{6} \ea
\ee  
and the concurrence of this state is given by 
\be 
\label{y24} 
C_{ia}=\frac{\eta}{3}\,\mbox{Max}\,\left[0,\,(4x-1)- \sqrt{2(1-x)(3-(1+2x)\eta^2)} \right]. 
\ee 
It is not difficult to observe that for all values of $x<0.625$, sudden death of entanglement occurs while for $x\geq 0.625$, there is an exponential decay of entanglement. 

The action of phase noise on the first qubit of the isotropic state leads to
\be 
\label{y25}
\rho_{ip} = 
 \frac{1}{6}\ba{cccc}
2(1-x) & 0 & 0 & 0 \\  0 & 2x+1    &  (4x-1)  \gamma &  0 \\ 
0 &   (4x-1)  \gamma &  2x+1 & 0  \\ 0 & 0 & 0 & 2(1-x)
\ea. \ee 
and 
\be 
\label{y26}
C_{ip}=\frac{1}{3}\,\mbox{Max} \left[0,(4x-1)\gamma-2(1-x) \right].
\ee  
Except at $x=1$, where we have an exponential decay of entanglement, one can see a finite time disentanglement for all values of $x$ in the open interval $\left(\frac{1}{2},\,1\right)$. 

When we allow the first qubit of the isotropic state to interact with depolarizing noise, the time evolved density matrix 
$\rho_{id}=\sum_{i=1}^4 \, K_{id}\,\rho\,K_{id}^\dagger$ retains its X form (see Eq.(1)) with 
\begin{eqnarray}
\label{y27}
a&=&d=\frac{3-3x+p(4x-1)}{9}; z=\frac{(3-4p)(4x-1)}{18} \nonumber \\
b&=&c=\frac{3+6x+2p(1-4x)}{18}; 
\end{eqnarray}
A simplified expression for concurrence is given by
\be 
\label{y28} 
C _{id}=\frac{1}{3}\,\mbox{Max}\,\left[0,\, 2p(1-4x)+6x-3  \right]. 
\ee  
With $p=1-e^{-\Gamma_d\,t/2}$, it can be seen that sudden death of entanglement occurs here for all values of $x>\frac{1}{2}$. 

\noindent{\bf Werner State}: Werner states are $d^2$ dimensional bipartite states that remain invariant under unitary transformations of the form $U\times U$~\cite{wer13}. A two-qubit Werner state is a mixture of the fully mixed state $\frac{I}{4}$ with probability $(1-x)$  and a singlet state $ \vert \psi \rangle =\frac{\vert 01 \rangle-\vert 10 \rangle}{\sqrt{2}}$ with  probability $x$~\cite{bruss}. That is, 
\be 
\rho_w = (1-x)\frac{I}{4} + x \vert \psi \rangle \langle \psi \vert.  
\ee 
As $ C_w =\mbox{Max}\, [0,\,\frac{3x-1}{2}]$, we have $C_w \neq 0 $ for all  $x > \frac{1}{3} $.

On allowing the first qubit of the Werner state to interact with amplitude noise we have
\be
\label{y30}
\rho_{wa} = 
 \ba{cccc}
\frac{(1-x)\eta^2}{4} & 0 & 0 & 0 \\  0 & \frac{(1+x)\eta^2}{4}  &  \frac{x\eta}{2}  &  0 \\ 
0 &   \frac{x\eta}{2} & \frac{2+(x-1)\eta^2}{4} & 0  \\ 0 & 0 & 0 & \frac{2-(1+x)\eta^2 }{4}\ea, 
\ee  
and 
\be 
\label{y31} 
C_{wa}=\frac{\eta}{2}\,\mbox{Max}  \left[0,\, 2x- \sqrt{(1-x)(2-(1+x)\eta^2)}  \right]. \ee 
It is not difficult to see that the action of amplitude noise on a single qubit of the two-qubit Werner state leads to sudden death of entanglement when the parameter $x$ lies in the range $\frac{1}{3}< x < \frac{1}{2}$. For $x\geq \frac{1}{2}$, there is an exponential decay of entanglement.

Subjecting the first qubit of Werner state with phase noise we have
\be 
\label{y32}
\rho_{wp} = \ba{cccc}
\frac{1-x}{4} & 0 & 0 & 0 \\  0 & \frac{1+x}{4}    &  \frac{x \gamma}{2}   &  0 \\ 
0 &   \frac{x \gamma}{2} &  \frac{1+x}{4} & 0  \\ 0 & 0 & 0 & \frac{1-x}{4}
\ea, 
\ee  
and
\be 
\label{y33} 
C_{wp}=\frac{1}{2}\, \mbox{Max}\, \left[0,\,2x \gamma-(1-x)\right]. 
\ee  
Here, it can be seen that there is strictly asymptotic decay of entanglement only when the parameter $x=1$. Though there is a finite time disentanglement for other values of $x$ i.e.,  $\frac{1}{3}<x<1$, the decay of entanglement is expected to be smooth and not very sudden.  

The action of depolarizing noise on the first qubit of the Werner state leads to
\be 
\label{y34}
\rho_{wd} = 
\ba{cccc}
\frac{3+(4p-3)x}{12} & 0 & 0 & 0 \\  0 & \frac{3+(3-4p)x}{12} &  \frac{(3-4p)x}{6} &  0 \\ 
0 &    \frac{(3-4p)x}{6} & \frac{3+(3-4p)x}{12} & 0  \\ 0 & 0 & 0 & \frac{3+(4p-3)x}{12}
\ea \ee 
and the concurrence of this state is given by
\be 
\label{y35} 
C_{wd}=\frac{1}{6}\,\mbox{Max}\, \left[0,\, 2(3-4p)x -(3+(4p-3)x) \right] 
\ee  
Here too, sudden death of entanglement is seen to occur for all values of $x$ in the range $\frac{1}{3}<x\leq 1$ as in the case of isotropic state.
\section{Conclusion} 
\label{sec4}
In this article, we have shown that a single noise acting on any one of the qubits of an entangled two-qubit state can cause finite time disentanglement. We have illustrated this fact by subjecting one of the qubits of a class of two-qubit X states and two-qubit pure states individually to three different kind of noises: amplitude, dephasing and depolarizing. While single-qubit channels of amplitude and phase noises are shown to be ineffective in causing finite time disentanglement in entangled pure states, single-qubit depolarizing noise is shown to cause sudden death of entanglement even in pure states. 

It is of importance to note here that the `mixedness' of a two-qubit state implies it might have undergone some kind of decoherence in the past, prior to the onset of any noise that we have considered. Thus, even when we have subjected only one qubit of the two-qubit mixed state to a single noise, the resultant sudden death of entanglement cannot conclusively be attributed to this applied noise alone. 
A pure state, on the other hand, is free from any {\emph {apriori}} decoherence and therefore, the sudden death of entanglement caused due to single noise acting on single qubit of any two-qubit pure state is due to the applied noise alone. Thus, our observation of sudden death of entanglement  in two-qubit pure states caused due to single-qubit depolarizing noise confirms that \emph{composite noise, in general, is not necessary for finite time disentanglement}. While not all single-qubit noises  are effective in causing sudden death of entanglement in two-qubit pure states, our analysis of depolarzing noise points at the possibility of existence of noisy environments that can cause finite time disentanglemet when acting alone on a single qubit of the two-qubit state. It can therefore be concluded that entanglement sudden death (ESD) is a natural phenomena which can happen even when only one part of an entangled system is exposed to an {\em appropriate} noisy environment.  

We wish to emphasize here that while the possibility of finite time disentanglement for two qubit X states under the action of {\em classical dephasing noise} on only one of the qubits is pointed out by Yu and Eberly in \cite{yucl}, the occurence of ESD due to the application of 
{\em local quantum noise} on a qubit of two-qubit entangled states has not been given due atention. 
We hope our work will initiate discussion in this regard.

\section*{Acknowledgement:}  
We thank the referee for the insightful suggestions in the light of which the article is revised. 
Yashodamma K.O acknowledges the support of Department of Science and Technology (DST), Govt. of India through the award of INSPIRE fellowship.

\end{document}